\documentclass{ws-ijgmmp}

\usepackage{xcolor}
\usepackage{float}
\usepackage[verbose,hypertexnames=false]{hyperref}
\hypersetup{colorlinks=false,allbordercolors=blue,pdfborderstyle={/S/U/W 1}}
\begin{document}

\markboth{Noraiz Tahir, Mubasher Jamil, Kaynat Fatima, and Tajammal Hussain Khookhar}
{Estimating the black hole shadow parameters and quasi-normal modes for Weyl-incorporated gravity}

%
\catchline{}{}{}{}{}
%

\title{Black hole shadow parameters and quasi-normal modes for Weyl-incorporated gravity}

\author{Noraiz Tahir\footnote{noraiz.tahir@sns.nust.edu.pk}}

\address{Department of Physics and Astronomy, School of Natural Sciences (SNS), National University of Sciences and Technology (NUST), 44000, Islamabad, Pakistan.\\
\email{noraiz.tahir@sns.nust.edu.pk}}

\author{Mubasher Jamil}

\address{Department of Mathematics, School of Natural Sciences (SNS), National University of Sciences and Technology (NUST), 44000, Islamabad, Pakistan. \\
\email{mjamil@sns.nust.edu.pk}}

\author{Kaynat Fatima}

\address{Department of Mathematics, School of Natural Sciences (SNS), National University of Sciences and Technology (NUST), 44000, Islamabad, Pakistan. \\
\email{fatimakaynat09@gmail.com}}
	
\author{Tajammal Hussain Khokhar}
\address{Department of Physics and Astronomy, School of Natural Sciences (SNS), National University of Sciences and Technology (NUST), 44000, Islamabad, Pakistan. \\
\email{tajammal.khokhar@sns.nust.edu.pk}}

\maketitle

\begin{history}
\received{(26 Feb 2026)}
\accepted{(15 July 2026)}
\end{history}

\begin{abstract}
An additional term of the form $\lambda \mathbf{T} \cdot \mathbf{C} \cdot \mathbf{T}$ in the Einstein-Hilbert Lagrangian was introduced to explain the interaction between matter and pure gravitational field [H. W. Lee and A. Qadir, Motion of test particle for Weyl-interaction gravity, {\it Int. Jour. Mod. Phys. D} {\bf 28}(16) (2019) 2040014], the modified relativistic dynamics (MORD). In this paper, we estimate the shadow parameters of black hole in a spherically symmetric static spacetime within the MORD framework. As a first approximation, we assume that the black hole is surrounded by a constant‑density baryonic matter halo. The analysis is then extended by introducing a homogeneous plasma background. We compute the shadow radii for various black hole masses and analyze their dependence on the WIG coupling parameter $\lambda$. In addition, we compute the fundamental quasi‑normal mode (QNM) frequencies  under test‑field approximation, and perform a time‑domain integration analysis.
\end{abstract}



\section{Introduction \label{intro}}
The landmark observations of M87* and Sgr A* by the EHT collaboration have inaugurated a new era in strong-field gravitational physics \cite{akiyama2019first, akiyama2022first}. These images provide direct visual confirmation of the black hole shadow whose size and morphology are governed by the geometry of spacetime near the event horizon \cite{michell1784vii, laplacepierre, einstein1922general, oppenheimer1939continued, johnwheeler, hawking1970singularities}. The black hole shadow therefore serves as a powerful probe of strong-field gravity, motivating theoretical studies of shadow formation in various gravitational frameworks.

A wide range of modified gravity theories has been proposed, and substantial effort has been devoted to check how such modifications affect black hole spacetimes and their observable shadows (c.f. \cite{Bambi2019, Johannsen2016, Psaltis2015, kruglov2020shadow, Bambi2012, Wang2021, saira, khadija, jusufi, k.jafarzade}). In most cases, these frameworks introduce additional degrees of freedom or free parameters to fit with observational data. While this flexibility enables a rich spectrum of phenomenological behaviors, it can also lead to increased model complexity and degeneracies among parameters, which may complicate the interpretation of observational constraints. While no single approach is preferred a \textit{priori}, complementary strategies that emphasize reduced parameter spaces remain of particular interest in the era of high-precision black hole observations. This perspective motivates the study of comparatively \textit{minimal} extensions of GR, which seek to encode possible deviations from standard gravity through a small number of well-motivated couplings, thereby facilitating clearer theoretical interpretation.

This work focuses on one such proposal, MORD, introduced originally in Refs. \cite{2017, lee2019motion}; here we refer to it as ``Weyl-incorporated gravity (WIG)''. The theoretical motivation for WIG originates from an analogy with quantum field theory (QFT). In the Einstein-Hilbert action, the matter sector $\sqrt{-g}\,T$ couples to gravity only through the metric tensor and the volume element; there is no explicit interaction between the stress–energy tensor $T^{\mu \nu}$ and the free radiative part of the gravitational field. In electromagnetism, by contrast, the field strength tensor $F_{\mu \nu}$ couples directly to the spin of particles. The Weyl tensor $C_{\alpha\mu\beta\nu}$ is the trace-free, conformally invariant part of the Riemann curvature and is the purest representation of a pure gravitational field in GR, analogous to $F_{\mu \nu}$ in electromagnetism \cite{lee2019motion, hofmann2013, jwheeler2018}. This analogy motivated the authors in Ref. \cite{lee2019motion} to introduce an interaction term $\lambda C_{\alpha\mu\beta\nu}T^{\alpha\beta}T^{\mu\nu}$ in the standard Einstein-Hilbert Lagrangian, where $\lambda$ is the coupling constant describing the interaction of matter with a pure gravitational field.

It was proposed that \textit{if} WIG is truly a minimal extension to GR, the new coupling constant $\lambda$ should be unique. The original proposed test was to check whether the WIG framework can explain the flatness of the rotational velocity curves of galaxies, thereby replacing the need for exotic dark matter. Recently, for a sample of eight edge-on spiral galaxies in the local group, a single value of the coupling constant, $\lambda = 4.3125 \pm 0.0012$, was shown to account for the asymptotic rotational velocities \cite{qadir2025}.

In this paper, we perform a theoretical exploration of the WIG framework by computing black hole shadow parameters and QNM frequencies. Specifically, we estimate the shadow radius for spherically symmetric static black holes embedded in a constant-density baryonic matter halo, and we analyze the dependence of the shadow on the coupling parameter $\lambda$. We also compute the fundamental QNM frequencies of massive scalar perturbations as a probe of the stability properties of the spacetime. Our aim is to understand the phenomenology of WIG in the strong-field regime and to explore whether the coupling parameter exhibits any dependence on the black hole mass scale. We adopt geometric units with $G = c = 1$.

The paper is arranged as follows: in Section~\ref{photonsphere} we estimate the black hole shadow parameters in the WIG framework, deriving the lapse function, effective potential, and shadow radius. In Section~\ref{energyemissionsection} we estimate the energy emission of the black holes for the obtained values of $\lambda$. In Section~\ref{plasmashadows} we extend the analysis with a homogeneous plasma background and estimate the variation in the shadow parameters. In Section~\ref{qnms} we compute the quasi-normal mode frequencies of massive scalar perturbations on the WIG background using the Wentzel-Kramers-Brillouin (WKB) approximation and the Asymptotic Iteration Method (AIM). We also study the time-domain evolution of these scalar perturbations in Section~\ref{timedomainintegration}. Lastly, in Section~\ref{results} the results are discussed.

\section{Black Hole Solution Embedded in a Baryonic Matter Halo in WIG \label{photonsphere}}
The WIG Lagrangian is given as \cite{lee2019motion}
\begin{align}
	\mathcal{L} = \sqrt{-g} \left( R - 2\Lambda - \kappa T + \lambda C_{\alpha\mu\beta\nu} T^{\alpha\beta} T^{\mu\nu} \right),
	\label{wiglagrangian}
\end{align}
where $\Lambda$ is the cosmological constant, and $\kappa = 8\pi$ is the gravitational coupling constant. By applying the variation on eq. (\ref{wiglagrangian}) we obtain the field equation as (see Section 3 and Appendix A of  Ref. \cite{lee2019motion} for detailed derivation)
\begin{align}
R_{\mu\nu} - \frac{1}{2} g_{\mu\nu} R + g_{\mu\nu} \Lambda = \kappa T_{\mu\nu} + \lambda I_{\mu\nu}, 
\label{variation}
\end{align}
where, $I_{\mu\nu}$ is the interaction term and is given by \cite{lee2019motion}
\begin{align}
I_{\mu\nu} = &\ \frac{1}{4} \Big( 
- g_{\alpha\beta} g_{\rho\mu} g_{\sigma\nu}
- g_{\rho\sigma} g_{\alpha\mu} g_{\beta\nu}
- g_{\alpha\sigma} g_{\rho\mu} g_{\beta\nu}\ 
- g_{\rho\beta} g_{\alpha\mu} g_{\sigma\nu} 
- g_{\beta\sigma} g_{\alpha\mu} g_{\rho\nu} 
\Big)\nonumber \\ & \Box T^{\alpha\beta} T^{\rho\sigma} + \frac{1}{6} \left( g_{\alpha\beta} g_{\rho\sigma} - g_{\rho\beta} g_{\sigma\alpha} \right)
\left( g_{\mu\nu} \Box - \nabla_\mu \nabla_\nu \right) T^{\alpha\beta} T^{\rho\sigma}.
\label{I_mu_nu}
\end{align}
Following Ref. \cite{lee2019motion} we assume a spherically symmetric static metric  ansatz as
\begin{align}
	ds^2 = -f(r) \, dt^2 + \frac{1}{f(r)} \, dr^2 + r^2 d\Omega^2,
		\label{shawrzchildwithf}
\end{align}
where, $f(r)$ is the lapse function. Also, we assume a pressureless dust distribution and the energy-momentum tensor $T_{\mu \nu}$ is then given by
	\begin{align}
		T_{\mu\nu} = \rho \, u_\mu u_\nu,
	\end{align}
	where $u^\mu = (1/\sqrt{f(r)}, 0, 0, 0)$ is the four-velocity of the static matter. The non-zero components are
	\begin{align}
		T_{tt} = \rho \, f(r), \qquad T_{rr} = T_{\theta\theta} = T_{\phi\phi} = 0,
	\end{align}
and the trace is $T = g^{\mu\nu} T_{\mu\nu} = -\rho$. Eq. (\ref{I_mu_nu}) represents the additional interaction contribution arising from the variation of the Weyl-matter coupling term in eq. (\ref{wiglagrangian}). Physically, it encapsulates the dynamical feedback between matter and the free, radiative part of the gravitational field. Unlike in GR, where the geometry responds linearly to the local energy-momentum distribution, the presence of $I_{\mu\nu}$ introduces a nonlinear coupling between matter anisotropies and tidal gravitational degrees of freedom. This term effectively modifies the spacetime curvature in regions of strong gravity or large matter gradients, acting as a correction to the conventional stress-energy tensor. In the weak-field limit ($\lambda \rho \to 0$), $I_{\mu\nu}$ vanishes, ensuring a smooth recovery of Einstein's theory \cite{lee2019motion}. We would like to emphasize that unlike in standard GR, $T_{\mu\nu}$ in eq.~\eqref{variation} is \emph{not} individually conserved. Using the explicit form of $I_{\mu\nu}$ given in eq.~\eqref{I_mu_nu}, the covariant divergence $\nabla^\mu I_{\mu\nu}$ evaluates to
 \begin{align}
 	\nabla^\mu I_{\mu\nu} = &\ \frac{1}{4} \mathcal{A}^{\alpha\beta\rho\sigma}_{\mu\nu} \left( \nabla^\mu \Box T^{\alpha\beta} \right) T^{\rho\sigma} 
 	+ \frac{1}{4} \mathcal{A}^{\alpha\beta\rho\sigma}_{\mu\nu} \Box T^{\alpha\beta} \left( \nabla^\mu T^{\rho\sigma} \right) \nonumber \\
 	&+ \frac{1}{6} \mathcal{B}_{\alpha\beta\rho\sigma} \left[ - R^\sigma_{\ \nu} \nabla_\sigma T^{\alpha\beta} T^{\rho\sigma} 
 	+ \mathcal{C}_{\mu\nu} \left( \nabla^\mu T^{\alpha\beta} \right) T^{\rho\sigma} 
 	+ \mathcal{C}_{\mu\nu} T^{\alpha\beta} \left( \nabla^\mu T^{\rho\sigma} \right) \right],
 	\label{divI_explicit}
 \end{align}
 where
 \begin{align}
 \mathcal{C}_{\mu\nu} &= g_{\mu\nu} \Box - \nabla_\mu \nabla_\nu, \\
 \mathcal{A}^{\alpha\beta\rho\sigma}_{\mu\nu} &= - g_{\alpha\beta} g_{\rho\mu} g_{\sigma\nu} - g_{\rho\sigma} g_{\alpha\mu} g_{\beta\nu} - g_{\alpha\sigma} g_{\rho\mu} g_{\beta\nu} - g_{\rho\beta} g_{\alpha\mu} g_{\sigma\nu} - g_{\beta\sigma} g_{\alpha\mu} g_{\rho\nu}, \\
\mathcal{B}_{\alpha\beta\rho\sigma} &= g_{\alpha\beta} g_{\rho\sigma} - g_{\rho\beta} g_{\sigma\alpha}.
 \end{align}
 One can now write the conservation equation as 
\begin{align}
	\nabla^\mu T_{\mu\nu} = -\frac{\lambda}{\kappa} \nabla^\mu I_{\mu\nu}.
\end{align}
Thus, $T_{\mu\nu}$ is not individually conserved; instead, there is an energy-momentum exchange between matter and the Weyl interaction sector, governed by the coupling $\lambda$. Thus, the non-minimal Weyl-matter coupling $ \lambda C_{\alpha\mu\beta\nu} T^{\alpha\beta} T^{\mu\nu} $ in the Lagrangian leads to an energy-momentum exchange between the matter and the gravitational interaction sector. This is a known feature of modified gravity theories with derivative or curvature-matter couplings \cite{Bambi2019, Johannsen2016, Psaltis2015, kruglov2020shadow, Bambi2012, Wang2021, saira, khadija, jusufi, k.jafarzade}. Nevertheless, the total energy-momentum, which includes the matter and the interaction terms, remains conserved, ensuring consistency with diffeomorphism invariance.

The components of $I_{\mu \nu}$ can be written as \cite{lee2019motion}
\begin{align}
	& I_{tt} = -\frac{1}{2r} \left[ r \left(\frac{f'(r)}{f(r)}\right)^2 \rho^2  - 2r \left(\frac{f'(r)}{f(r)}\right) \rho \rho' - 4r \rho'^2  - 4r \rho \rho''  \right. \nonumber \\ &  \left. + 2r \left(\frac{f(r)}{f'(r)}\right) \rho \rho' - 8 \rho \rho' \right],
	\label{Itt}
\end{align}

\begin{equation}
	I_{rr} = \frac{1}{2} \left(\frac{\rho \, f(r)}{f'(r)}\right)^2,
	\label{Irr}
\end{equation}
and 
\begin{align}
I_{\theta\theta} = I_{\phi\phi} = 0,
    \label{Ithetatheta}
\end{align}
respectively. Here $\rho$ is the density distribution of the dark matter surrounding the black hole. Note that when $\rho = 0$, we have $T_{\mu\nu}=0$ and $I_{\mu\nu}=0$, so eq.~\eqref{variation} reduces to $G_{\mu\nu} + g_{\mu\nu}\Lambda = 0$. Using the metric ansatz eq.~\eqref{shawrzchildwithf}, the $tt$ component yields
	\begin{align}
		\frac{1}{r^2}\left( r f'(r) + f(r) - 1 \right) + \Lambda = 0.
	\end{align}
Solving the above differential equation gives ${\displaystyle f(r) = 1 - \frac{2m}{r} - \frac{\Lambda}{3}r^2}$, which is the Schwarzschild–de~Sitter solution, and for $\Lambda=0$, we recover the ordinary Schwarzschild black hole solution. Thus, the WIG theory has a consistent vacuum limit.

Now the observations of the cosmic microwave background (CMB) observations indicate that $\simeq 5\%$ of the Universe should be made up of baryons (the usual protons and neutrons), but observations of the luminous parts of the galaxies show only half of these baryons, this is the ``missing baryon problem''. Since the baryons are dark, we call this the ``baryonic dark matter''. It is proposed that a significant fraction of this baryonic dark matter is present in the dark matter halos \cite{planck2016baryons, planck2020, gunn1996, gerhard1996, frasermcKelvie2011}. We assume that the full fraction of the dark matter halo surrounding the black hole contains these missing baryons. Following Refs. \cite{lee2019motion, qadir2025} One can write the Weyl-incorporated field equations (WIFE) as
\begin{align}
	G_{tt} = \frac{1}{r^2} \left( r \frac{f(r)}{f'(r)} + \frac{1}{f(r)} - 1 \right) = 8\pi f(r) \rho + \lambda I_{tt},
	\label{gtt}
\end{align}
and 
\begin{equation}
   G_{rr} = -\frac{1}{r^2} \left( -r \frac{f'(r)}{f(r)} + \frac{1}{f(r)} - 1 \right) =  \lambda I_{rr}.
   \label{grr}
\end{equation}
Solving eqs. (\ref{gtt}) and (\ref{grr}) \textbf{simultaneously}, we get the lapse function as
\begin{align}
    f(r)  =  \frac{\exp(-F(r)) \, r}{1 - \lambda r \rho \rho'} \left( \int_{r_0}^r \frac{\exp(F(q))}{q^2}\, dq - \int_{r_0}^r \rho \, \exp(F(q)) \,dq \right),
    \label{mu*}
\end{align}
where $F(r)$ is
\begin{align}
F(r) = \frac{1}{2} \int_{r_0}^r \frac{4 + \lambda q \, \mathcal{E}(q)}
{q(1 - \lambda \, q \, \rho \rho')}\, dq, 
\label{F(r)}
\end{align}
\begin{align}
    \mathcal{E} (r) = r \rho^2 \left(\frac{f'(r)}{f(r)}\right)^2 - 2 r \rho \rho' \left(\frac{f'(r)}{f(r)}\right) - 2 r \rho'^2 - 2 r \rho \rho'' - 8 \rho \rho'.
    \label{mathcalE}
\end{align}

\subsection{A Constant Density Case of Baryonic Halo}
\label{constantdensity}
As a first approximation, we consider a black hole embedded in a dark matter halo with constant density, $\rho \rightarrow \rho_c$. We would like to point out that our present focus is on isolating and understanding the fundamental modifications introduced by the coupling parameter $\lambda$. Treating the surrounding dark matter halo as a homogeneous background provides a valuable pedagogical framework for disentangling the effects of the coupling while avoiding additional model-dependent complexities.

Before proceeding, we clarify the scope and limitations of our metric ansatz. The form in eq. (\ref{shawrzchildwithf}) implicitly assumes $G^t_{\,t} = G^r_{\,r}$. For a pressureless matter distribution $T^t_{\,t} \neq T^r_{\,r}$ and the $I_{\mu\nu}$ 
term which generally gives $I^t_{\,t} \neq I^r_{\,r}$ this ansatz is not exact. However, for a constant-density halo and sufficiently small 
coupling $\lambda$, the deviations from equality are expected to be small. We therefore 
treat this as a leading-order approximation to isolate the main effects of the Weyl 
coupling. A fully consistent treatment using the more general metric 
$ds^2 = -A(r)dt^2 + dr^2/B(r) + r^2 d\Omega^2$ with $A(r) \neq B(r)$, along with a 
non-homogeneous baryonic dark matter halo, is deferred to future work.

Under the constant density assumption eqs. (\ref{mu*}) - (\ref{mathcalE}) can be reduced to
\begin{equation}
    f(r) = \exp(-F_c(r)) \, r \left(\int_{r_0}^r \frac{\exp(F_c(q))}{q^2}\, dq - \rho_c \int_{r_0}^r \exp(F_c(q)) \, dq \right), 
    \label{constantlapse}
\end{equation}
where, 
\begin{equation}
    F_c(r)= 2 \ln \left(\frac{r}{r_0}\right)  + \int_{r_0}^r \lambda \mathcal{E}_c(q) \, dq, 
    \label{constantF}
\end{equation}
and 
\begin{equation}
   \mathcal{E}_c(r) = r \rho_c^2 \left(\frac{f'(r)}{f(r)}\right)^2.
    \label{constantalpha}
\end{equation}
To obtain the explicit behavior of photons around the black hole, we use an order-by-order approximation. For the zeroth-order approximation, we consider that $\lambda\rho_c \rightarrow 0$, and hence from eqs. (\ref{constantF}) and (\ref{constantalpha}) we get
\begin{equation}
    F_c^{(0)}(r)= 2 \ln \left(\frac{r}{r_0}\right),
    \label{constantFzero}
\end{equation}
and 
\begin{equation}
    \mathcal{E}^{(0)}(r)= r \rho_c \left(\frac{f'_{(0)}(r)}{f_{(0)}(r)}\right)^2.
    \label{alphazero}
\end{equation}
From eq. (\ref{constantlapse}) the zeroth-order lapse function will be
{\bf \begin{equation}
    f_{(0)}(r) = 1- \frac{2m}{r}.
    \label{zerolapseconstant}
\end{equation}}
This is the same solution of the standard Einstein's field equation when $\lambda = 0$ in eq. (\ref{variation}). Now to find the first-order correction term, eq. (\ref{constantF}) can be written as 
\begin{align}
    F_c^{(1)} (r) & = 2 \ln \left(\frac{r}{r_0}\right) + \frac{\lambda}{2} \int_{r_0}^r \mathcal{E}(q)\, dq.
    \label{constantF1st}
\end{align}
As a result the first-order corrected lapse function will become
\begin{align}
    f_{(1)}(r) = & \exp(-F_c^{(1)}(r)) \left[r \left(\int_{r_0}^r \frac{\exp(F_c^{(1)}(q))}{q^2}\, dq  - 8\pi\rho_c \int_{r_0}^r \exp(F_c^{(1)}(q)) \, dq \right)\right]. 
    \label{first}
\end{align}
The horizon structure of a black hole is determined by the zeros of the lapse function, which correspond to event horizons where the gravitational redshift becomes infinite (see \cite{HawkingEllis1973, Penrose1965, Ashtekar2004, Senovilla2011, Visser1998} for detailed discussions). The causal structure of the spacetime is determined by the sign of the lapse function, and in our case, the sign of $f_{(1)}(r)$. For regions where $f_{(1)}(r) > 0$, the $t$-coordinate is timelike, and $r$ is spacelike, corresponding to the exterior of the black hole. Conversely, when $f_{(1)}(r) < 0$, the roles of $t$ and $r$ interchange, and the metric describes the interior region where no static observers can exist. The hypersurface defined by $f_{(1)}(r) = 0$ marks the transition between these two causal domains: it is a null surface where $g_{tt} = 0$ and $g_{rr} = 0$, implying that light rays directed radially outward remain trapped. Consequently, this surface corresponds to an event horizon, where the gravitational redshift of outgoing radiation diverges. The zeros of $f_{(1)}(r)$ therefore determine the location of the black hole horizons. 

Before presenting our numerical results, here we can derive the expressions to quantify the error introduced by the metric ansatz. For a general spherically symmetric metric the difference between the Einstein tensor components is
\begin{equation}
G^t_{\;t} - G^r_{\;r} = \frac{B}{r}\left(\frac{A'}{A} - \frac{B'}{B}\right).
\label{eq:gr_diff}
\end{equation}
In our ansatz since $A(r)=B(r)=f(r)$, so the left-hand side vanishes identically. From the field equations, the exact relation is
\begin{equation}
G^t_{\;t} - G^r_{\;r} = \kappa (T^t_{\;t} - T^r_{\;r}) + \lambda (I^t_{\;t} - I^r_{\;r}).
\label{eq:exact}
\end{equation}
For pressureless dust, since $T^t_{\;t} = -\rho$ and $T^r_{\;r} = 0$, and for constant density eqs.~\eqref{Itt}-\eqref{Irr} give $I^t_{\;t} - I^r_{\;r} = \mathcal{O}(\rho_c^2)$. Thus
\begin{equation}
G^t_{\;t} - G^r_{\;r} = -\kappa \rho_c + \lambda \mathcal{O}(\rho_c^2).
\label{eq:deviation}
\end{equation}
The relative error in our ansatz is therefore
\begin{equation}
\epsilon \equiv \frac{|G^t_{\;t} - G^r_{\;r}|}{|G^t_{\;t}|} \equiv \rho_c r^2,
\label{eq:epsilon}
\end{equation}
where we have used $|G^t_{\;t}| \sim 1/r^2$ from the Schwarzschild contribution. The perturbative expansion is controlled by the small parameter
\begin{equation}
\delta \equiv \lambda \rho_c.
\label{eq:delta}
\end{equation}

\begin{figure}
    \centering
    \includegraphics[width=0.9\linewidth]{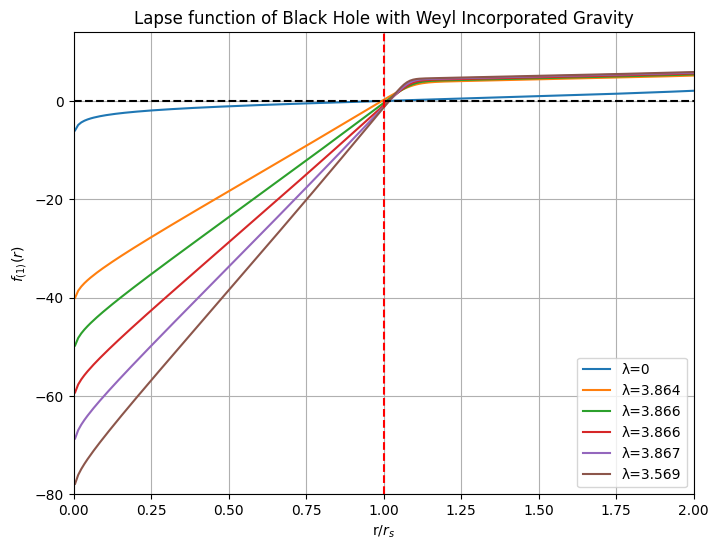}
    \caption{The lapse function for spherically symmetric static black holes with different masses $m$, and corresponding $\lambda$.}
    \label{lapsefuntionprofile}
\end{figure}

\begin{figure}
	\centering
	\includegraphics[width=1.0\linewidth]{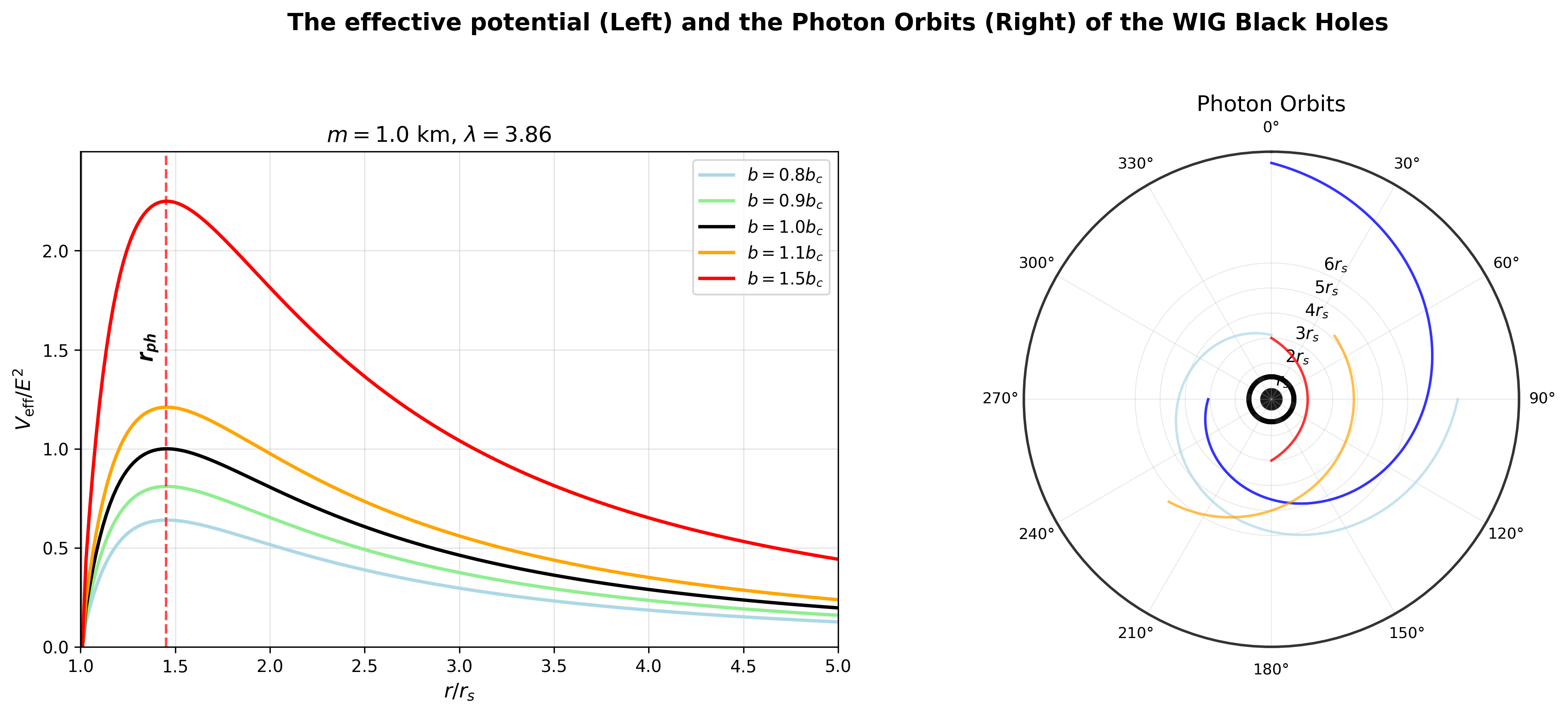}
    \caption{Effective potential profiles (left panel) and corresponding photon trajectories (right panel) for WIG black hole with $m= 1$ km and $\lambda = 3.869$. Left panel: $V_{\text{eff}}(r)/E^2$ versus $r/r_s$ for impact parameters $b = 0.8b_c, 0.9b_c, b_c, 1.1b_c, 1.5b_c$ (color-coded). The photon sphere ($r_{\text{ph}}$, red dashed) and event horizon ($r_s$, black solid) are marked. Right panel: Polar representation of photon orbits showing captured ($b < b_c$, blue), critical ($b = b_c$, black), and deflected ($b > b_c$, red/orange) trajectories within circular boundary at $12r_s$. Angular coordinates ($0^0$–$360^0$) and radial markers ($r_s$–$6r_s$) provide scale reference. The value of $\mathcal{K} = 1$. \label{effective}}
\end{figure}

To investigate how $\lambda$ affects black hole solutions across different mass scales, we solve eq. (\ref{constantlapse}) for various astrophysical black hole masses. Specifically, we determine the values of $\lambda$ that satisfy the horizon condition $f_{(1)}(r_s) = 0$ at the Schwarzschild radius $r_s$.  Table \ref{photonsphereandshadowradius} presents the obtained $\lambda$ values for black holes spanning several orders of magnitude in mass with a constant halo density $\rho_c = 0.5$. For each mass configuration, we find a unique value of $\lambda$ that preserves the horizon structure, consistent with expectations for spherically symmetric static geometries \cite{Schwarzschild1916, Penrose1969}. The corresponding lapse function profiles for different mass parameters and their associated $\lambda$ values are shown in Fig. \ref{lapsefuntionprofile}. One can see that for the larger $\lambda$ the horizon lies slightly inside $r=2m$. This inward shift is a genuine effect of the Weyl-matter coupling in the presence of a constant‑density halo. The non‑smooth appearance of the curves near $r=2m$ is an artifact of the constant‑density approximation and the metric ansatz. A realistic halo density that vanishes at the horizon would restore a smooth crossing.
To estimate the expression for $V_{\rm eff}$, we define the Lagrangian in terms of $f_{(1)}(r)$ as
\begin{align}
\mathcal{L} = \frac{1}{2} \left( - f_{(1)} (r) \dot{t}^2 + \frac{1}{f_{(1)} (r)} \dot{r}^2 + r^2 \dot{\theta}^2 + r^2 \sin^2\theta \dot{\phi}^2 \right),
\end{align}
where the dot represents the derivative with respect to $\tau$. From the Euler-Lagrange equation, we get 
\begin{align}
E= \frac{1}{2}\left( \frac{e^{F_{c}^{(1)}(r)}}{r \left( \int \frac{e^{F_{c}^{(1)}(r)}}{r^2} dr - \rho \int e^{F_{c}^{(1)}(r)} dr \right)}\right)\dot{t},
\label{testparticleenerygy}
\end{align}
and
\begin{align}
    L = r^2 \sin^2\theta \dot{\phi}.
    \label{angularmomentum}
\end{align}
Here, $E$ and $L$ represent the test particle energy, and angular momentum. Following the standard Carter separation procedure \cite{carter, hamil}
\begin{equation}
    V_{\rm eff}(r)=f_{(1)}(r) \left( \frac{L^2 + \mathcal{K}}{r^2} - E \right).
    \label{effectivepotentialequation}
\end{equation}
where, $\mathcal{K}$ is the Carter constant. To identify the unstable orbits, we impose the following condition
\begin{equation}
	V_{\text{eff}}(r) \bigg|_{r = r_p} = 0, \qquad 
	\frac{dV_{\text{eff}}(r)}{dr} \bigg|_{r = r_p} = 0, \qquad 
	\frac{d^2 V_{\text{eff}}(r)}{dr^2} \bigg|_{r = r_p} < 0.
	\label{27}
\end{equation}
where, $r_p$ denotes the radius of the photon sphere. Solving eq. (\ref{27}) for the first two conditions we get \footnote{The inequality $V''_{\text{eff}}(r_p) < 0$ (ensuring instability) holds for all solutions considered here, as verified numerically.}
\begin{align}
        r \frac{df_{(1)}(r)}{dr}\bigg|_{r = r_p} + 2 f_{(1)}(r)\bigg|_{r = r_p} = 0. 
        \label{numericaleq}
\end{align}
One can define the impact parameters as \cite{hamil}
    \begin{align}
        \zeta = \frac{L}{E},
        \label{zeta}
    \end{align}
and
    \begin{align}
        \eta = \frac{\mathcal{K}}{E^2}.
        \label{eta}
    \end{align}
Taking the square of eq. (\ref{eta}) and add the resulting equation to eq. (\ref{zeta}). We then substitute eqs. (\ref{testparticleenerygy})-(\ref{angularmomentum}) in the obtained expressions. The impact parameter for $\mathcal{K}=1$ is given by
\begin{align}
    R_s^2 = \eta + \zeta^2 = \frac{r_p^2}{f_{(1)}(r_p)}. \label{zetaetea}
\end{align}
To determine the black hole shadow size, characterized by the impact parameter $\eta + \zeta^2$, one must first compute the photon sphere radius $r_p$. This is achieved by numerically solving eq.~(\ref{numericaleq}) for the photon orbit location. In Table~\ref{photonsphereandshadowradius} we give the resulting values of $r_p$ and the corresponding impact parameter for different mass configurations and their associated $\lambda$ values. Our calculations demonstrate a positive correlation between black hole mass and shadow size, with larger masses producing larger shadows. The shadow radii we obtain show consistency with values reported in Ref. \cite{hamil}.
\begin{table}
\tbl{The estimated black hole shadow parameters in WIG. The table shows photon radius $r_p$, impact parameter $\eta + \zeta^2$, and shadow radius $R_s$ for different masses $m$ and corresponding $\lambda$ values. \label{photonsphereandshadowradius}}
{\begin{tabular}{@{}ccccc@{}} \toprule
$m$ (km)  &$\lambda$ & $r_p$ (km) & $\eta+\zeta^2$ (km$^2$) & $R_s$ (km) \\ \colrule
$10^{-6}$ & $3.864$ & $0.355$ & $4.281$ & $2.069$\\
$10^{-5}$ &  $3.866$ & $0.535$ & $4.325$ & $2.079$\\
$10^{-2}$ & $3.867$ & $0.844$ & $4.712$ & $2.171$ \\
$1$ &  $3.869$  & $1.504$ & $5.234$ & $2.288$\\
$10^3$ & $3.568$ & $1.873$ & $5.847$ & $2.418$\\ \botrule
\end{tabular}}
\end{table}
In order to visualize the shape of black hole shadows, we use the celestial coordinates $X$ and $Y$ which yeild 
\begin{equation}
b = X^2 + Y^2 = \eta + \zeta^2 = R^2_s =  \frac{r_p^2}{f_{(1)}(r_p)}. \label{eqn:shadow_boundary}
\end{equation}
The effective potential for each $m =1$ km and the corresponding $\lambda$ is shown in Fig. 2 (Left Panel), for various values of the impact parameter, $b = R_s$ and the corresponding photon orbits are shown in the Right Panel of Fig. \ref{effective}. One can see that the effective potential reveals three distinct photon orbital regimes for black holes in WIG. Photons with $b < b_{c}$ (blue/green curves) encounter potential barriers that vanish near $r/r_s = 1$, leading to gravitational capture. The photon sphere manifests as a local maximum at $r_{\text{ph}}/r_s \approx 1.5\text{--}1.8$, where photons with critical impact parameter $b = b_{c}$ (black curve) execute unstable circular orbits. For $b > b_{c}$ (orange/red curves), centrifugal barriers cause photon deflection, with deflection angles decreasing as $R_s$ increases. The coupling constant $\lambda$ modifies $V_{\text{eff}}$ morphology, higher $\lambda$ values slightly elevate and shift the photon sphere outward relative to Schwarzschild, while $\lambda = 3.568$ for $m = 10^3$ km reduces barrier height by $\sim 7\%$, indicating weaker photon confinement. This trend indicates that the Weyl interaction effectively modifies the curvature of spacetime in the strong-field regime, leading to small but measurable deviations in the photon sphere and shadow size compared to the Schwarzschild limit. Importantly, across all mass configurations, the potential retains its qualitative Schwarzschild-like behavior, confirming that the fundamental causal structure of the spacetime remains preserved under WIG. 

Before proceeding, we clarify the interpretation of the $\lambda$ values in Table \ref{photonsphereandshadowradius}. From eqs. \eqref{eq:epsilon} and \eqref{eq:delta}, for $\rho_c = 0.5$, we have $\delta \equiv \lambda \rho_c \sim 1.9$ and $\epsilon \sim \rho_c r^2 \sim 0.5$ at $r \sim \mathcal{O}(1)$ km. Thus, the perturbative expansion is not controlled, confirming that our pedagogical choice probes the strong-coupling regime. These $\lambda$ values are obtained by solving the horizon condition for a black hole embedded in a constant-density halo. They are therefore {\it conditional} for the given specific halo model, the WIG field equations require $\lambda \sim 3.5$--$3.9$ to admit a black hole horizon. The value $\rho_c = 0.5$ is chosen to amplify the Weyl-matter interaction and isolate its effects. It is not intended to represent a realistic astrophysical halo density. The GR limit, $\lambda \rho_c=0$, remains a valid and distinct limit of the theory, corresponding to the standard Schwarzschild solution. Our analysis, therefore, explores a specific non-GR sector of the WIG framework and should not be interpreted as a constraint on $\lambda$ from observations.
\section{Energy Emission of Black Holes in WIG \label{energyemissionsection}}
\begin{figure}
\centering
\includegraphics[scale=0.5]{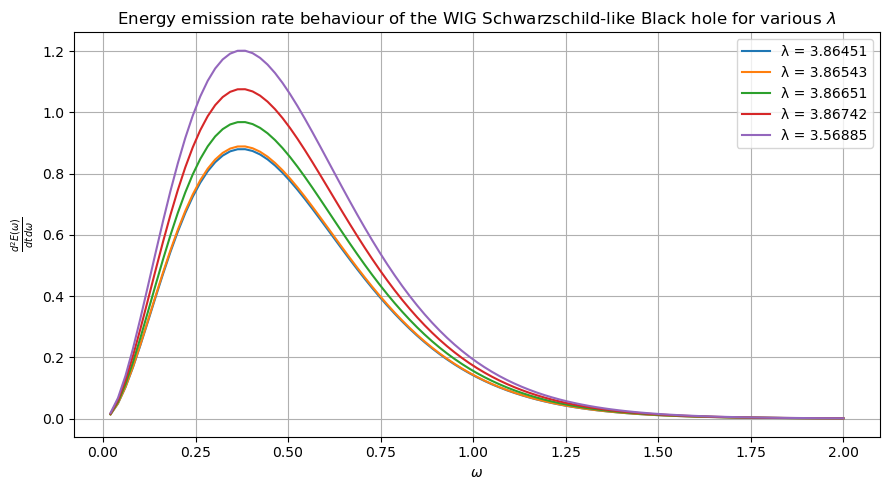}
\caption{The energy emission rates of the spherically symmetric static black holes with WIG. The variation in the energy emission rate is shown for various values of $m$ and corresponding $\lambda$. \label{energyemission}}
\end{figure}

A key area of investigation in black hole physics involves the study of their emission properties. This analysis focuses on the energy emission rate, a quantity that can be characterized using the geometric optics approximation. In this high-energy limit, the absorption cross-section of a black hole oscillates around a constant value, $\sigma_{\rm lim}$ and is defined as \cite{hamil}
\begin{align}
    \sigma_{\text{lim}} \sim \pi R_s^2.
\end{align}
The rate at which a black hole emits energy per unit time $t$ and per unit frequency $\omega$ is given by
\begin{align}
    \frac{d^2E(\omega)}{dt \, d\omega} = \frac{2\pi^2 \sigma_{\text{lim}} \, \omega^3}{e^{\omega/T} - 1},
\end{align}
where $T$ is the Hawking temperature, and $\omega$ is the frequency of the photon. Following the seminal work of Ref.~\cite{hawkingradiations}, the Hawking temperature of a Schwarzschild black hole of mass $m$ is $T = 1/(8\pi m)$. We adopt this as a leading-order approximation for our energy emission estimate. A full treatment would require computing the exact Hawking temperature $T = f'(r_h)/(4\pi)$ and the critical impact parameter $b_c$ from the metric $f(r)$ in eq.~(\ref{mu*}). However, due to the complexity of $f(r)$, we leave such a complete analysis for future work and use the Schwarzschild approximation here. In Fig. \ref{energyemission} we give the energy emission of the spherically symmetric static black hole with WIG for various values of $m$ and corresponding $\lambda$.
\section{Impact of Homogeneous Plasma Background on Shadow Parameters \label{plasmashadows}}
We assume a homogeneous plasma whose refractive index is $n(x^\mu, \omega)$, where $\omega$ is the photon frequency. Following Ref. \cite{synge1960} one can write the refractive index of a homogeneous plasma as
\begin{align}
    n(r) = \sqrt{1-\frac{\rho_p}{r}}.
    \label{plasmanumberdesnity}
\end{align}
where $\rho_p$ is the density of plasma background. The Hamiltonian Jacobi equation in the presence of the plasma background is given by \cite{synge1960}
\begin{equation}
    \frac{\partial \mathcal{S}} {\partial \tau} = -\frac{1}{2} \left( g^{\mu\nu} p_\mu p_\nu - ({n^2 - 1}){\left( {p_0} - g^{00} \right)^2} \right).
    \label{jacobiplasma}
\end{equation}
The new set of null geodesics become
\begin{align}
    \dot{t} = \frac{n^2 E}{f_{(1)}(r)},
\end{align}
\begin{align}
    L = r^2 \sin^2 \theta \, \dot\phi,
\end{align}
\begin{align}
    r^2 \dot{r} = \sqrt{r^4 n^2 E^2 - r^2 f_{(1)} (r) \left(L^2 + \mathcal{K}\right)},
\end{align}
and
\begin{align}
    r^2 \dot{\theta} = \sqrt{\mathcal{K} - L^2 \cot^2 \theta}.
\end{align}
The effective potential is given by
\begin{align}
    \tilde{V}_{\text{eff}}(r) =  \frac{r^2}{f_{(1)}(r)}(L^2+\mathcal{K}) - {n^2}{E^2}.
\end{align}
Using eq. (\ref{27}) we obtain the equation of the unstable orbit as 
\begin{equation}
\left(2r^4 n(r) \frac{dn(r)}{dr} E^2 + (L^2 + \mathcal{K}) \frac{df_{(1)}(r)}{dr}\right) \bigg|_{r = r_p} = 0.
    \label{unstableorbitwithplasma}
\end{equation}
The impact parameter $b$ is given by 
\begin{equation}
    \eta + \zeta^2 = r_p^2 \frac{n^2(r_p)}{f_{(1)}(r_p)}.
    \label{impactwithplasma}
\end{equation}
We use the same procedure followed in the previous sections to estimate the shadow radius. In Table \ref{parameterswithplasma} we give the estimated size of the photon sphere $r_p$ (column 3), the impact parameter $\eta + \zeta^2$ (column 4) and the corresponding shadow radius $R_s$ in column 5.
\begin{table}[ht]
\tbl{Black hole parameters in the presence of a homogeneous plasma background for different plasma densities $\rho_p$. Calculations are performed within the WIG framework.
\label{parameterswithplasma}}
{\begin{tabular}{@{}cccccccc@{}} \toprule
$m$ (km) & $\lambda$ & \multicolumn{3}{c}{$\rho_p=0.4$} & \multicolumn{3}{c}{$\rho_p=0.2$} \\
& & $r_p$ (km) & $\eta+\zeta^2$ (km$^2$)& $R_s$ (km) & $r_p$ (km) & $\eta+\zeta^2$ (km$^2$) & $R_s$ (km) \\ \colrule
$10^{-6}$ & $3.864$ & $1.945$ & $0.981$ & $0.991$ & $2.055$ & $1.899$ & $1.378$\\
$10^{-5}$ & $3.866$ & $1.785$ & $0.916$ & $0.957$ & $2.014$ & $1.852$ & $1.361$ \\
$10^{-2}$ & $3.867$ & $1.452$ & $0.898$ & $0.947$ & $1.925$ & $1.824$ & $1.350$\\
$1$ &  $3.867$  & $1.323$ & $0.877$ & $0.936$ & $1.895$ & $1.761$ & $1.327$ \\
$10^3$ & $3.568$ & $1.256$ & $0.847$ & $0.920$ & $1.752$ & $1.698$ & $1.303$\\ \botrule
\end{tabular}}
\end{table}
\begin{figure}[ht]
    \centering
    \includegraphics[width=.65\textwidth]{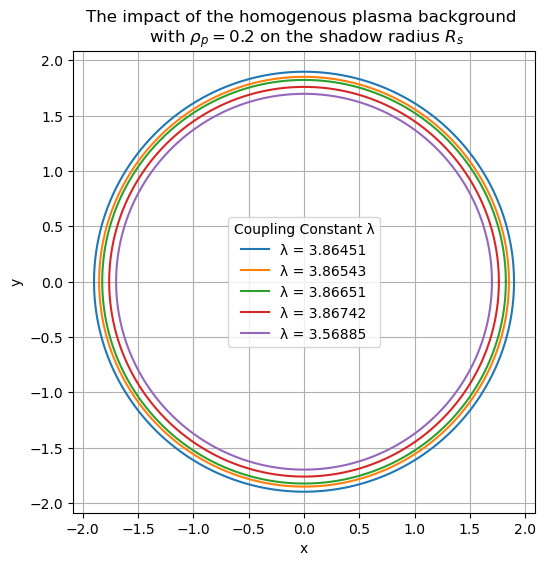}
    \qquad
    \includegraphics[width=.65\textwidth]{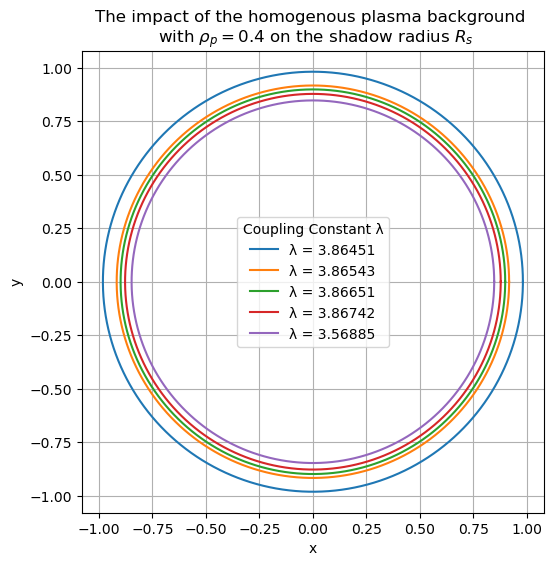}
    \caption{The impact of m and corresponding $\lambda$ on the impact parameter with a homogenous plasma background with $\rho_p=0.4$ (upper panel) and $\rho_p=0.2$ (lower panel).}%
    \label{plasmabackgroundimpact0.4and0.2}%
\end{figure}
\section{Quasi-Normal Modes and Mode Stability Analysis \label{qnms}}
The ringdown phase of a black hole, following a perturbation such as an in-falling particle or a gravitational wave, is characterized by the emission of QNMs. These complex-frequency oscillations ($\omega = \omega_R + i \omega_i$) provide a unique fingerprint of the black hole spacetime geometry, making them powerful probes for testing gravitational theories in the strong-field regime \cite{cheng2020qnms, jusufi2021qnms, ghasemi2020qnm}. 

We emphasize that the analysis presented in this section is performed in the test-field approximation. We perturb a massive scalar field on a fixed WIG background, while keeping the metric $g_{\mu\nu}$, the matter tensor $T_{\mu\nu}$, and the interaction term $I_{\mu\nu}$ unperturbed. A true gravitational QNM calculation in WIG would require perturbing all these fields simultaneously and solving the coupled system; this is a task beyond the scope of this work and will be considered separately. Our results are limited to the test-field approximation and the specific modes considered. A rigorous proof of mode stability would require a full perturbative analysis of the coupled metric-matter system, which is beyond the scope of this work. To ensure robustness of our test-field QNM calculations, we employ two independent semi-analytical methods: $(i)$ the WKB approximation; and $(ii)$ the AIM.

For a massive scalar field $\Phi$ on the WIG background, separation of variables $\Phi = (\psi_\ell(r)/r) Y_{\ell m}(\theta,\phi) e^{-i\omega t}$ yields the master equation \cite{Konoplya2003, Berti2009}
\begin{align}
	\frac{d^2 \psi_\ell}{dr_*^2} + \left[ \omega^2 - V_\ell(r) \right] \psi_\ell = 0,
	\label{masterequation}
\end{align}
where $dr_* = dr/f_{(1)}(r)$ is the tortoise coordinate and the effective potential is
\begin{align}
	V_{\ell}(r) = f_{(1)}(r) \left( \mu^2 + \frac{\ell(\ell+1)}{r^2} + \frac{1}{r} \frac{d f_{(1)}(r)}{dr} \right).
	\label{scalarpotentialmassive}
\end{align}

\subsection{The WKB Approximation Method}
We employ the 6th-order WKB approximation because it offers an optimal balance between computational tractability and accuracy for fundamental modes ($n=0$). For our focus on fundamental modes, the 6th-order WKB provides adequate precision to discern the effects of the coupling constant $\lambda$ on the QNM spectrum. The 6th-order WKB formula gives QNM frequencies as \cite{cheng2020qnms}
\begin{align}
\omega^2 = V_0 + \sqrt{-2 V_0''} \, \Lambda(n) - i \left( n + \frac{1}{2} \right) \sqrt{-2 V_0''} \left[ 1 + \Omega(n) \right],
\label{wkbformula}
\end{align}
where $V_0 = V_{\ell}(r_0)$ is the potential maximum at $r_0$, $V_0'' = \frac{d^2 V_{\ell}}{dr_*^2}\big|_{r=r_0}$, and $\Lambda(n)$ and $\Omega(n)$ are WKB correction terms.

\subsection{Asymptotic Iteration Method}
We begin with the master wave equation eq. (\ref{masterequation}). To apply the AIM, we first introduce the coordinate transformation
	\begin{align}
		\xi = 1 - \frac{r_s}{r}, \qquad \xi \in [0,1),
		\label{xicoord}
	\end{align}
	Under this transformation, the event horizon $r = r_s$ maps to $\xi = 0$, and spatial infinity $r \to \infty$ maps to $\xi \to 1$. We now express the wavefunction $\psi_\ell(r)$ in terms of $\xi$ using the ansatz
	\begin{align}
		\psi_\ell(r) = (\xi - 1)^\alpha \xi^\beta \chi_\ell(\xi),
		\label{wavefunctionansatz}
	\end{align}
	where the exponents $\alpha$ and $\beta$ are chosen to extract the correct asymptotic behavior near the horizon $\xi \to 0$ and at infinity $\xi \to 1$. Near the horizon, the effective potential vanishes and the master equation admits ingoing wave solutions, which fixes
	\begin{align}
		\alpha = -\frac{i \omega}{f'_{(1)}(r_s)}.
		\label{alpha}
	\end{align}
	At spatial infinity, for massive fields ($\mu \neq 0$), the asymptotic behavior gives
	\begin{align}
		\beta = \frac{1}{2} - \sqrt{\frac{1}{4} - \mu^2 r_s^2},
		\label{beta}
	\end{align}
	while for massless fields ($\mu = 0$), we have $\beta = 0$ up to an overall factor. Substituting the ansatz eq. (\ref{wavefunctionansatz}) into the radial equation and transforming from $r$ to $\xi$ yields a second-order differential equation for $\chi_\ell(\xi)$ of the form
	\begin{align}
		\chi_\ell''(\xi) = \mathcal{R}_0(\xi) \chi_\ell'(\xi) + s_0(\xi) \chi_\ell(\xi),
		\label{aimform}
	\end{align}
	where $\mathcal{R}_0(\xi)$ and $s_0(\xi)$ are functions that depend on $f_{(1)}(r(\xi))$, $\omega$, $\ell$, and $\mu$. The AIM is based on the observation that if $\mathcal{R}_0(\xi)$ and $s_0(\xi)$ satisfy the relation
	\begin{align}
		\frac{s_k(\xi)}{\mathcal{R}_k(\xi)} = \frac{s_{k-1}(\xi)}{\mathcal{R}_{k-1}(\xi)} \equiv \beta(\xi)
		\label{aimcondition}
	\end{align}
	for some $k$, then the solution $\chi_\ell(\xi)$ can be obtained in closed form. Here the sequences $\mathcal{R}_k(\xi)$ and $s_k(\xi)$ are generated by the recurrence relations
	\begin{align}
		\mathcal{R}_k(\xi) &= \mathcal{R}_{k-1}'(\xi) + s_{k-1}(\xi) + \mathcal{R}_0(\xi) \mathcal{R}_{k-1}(\xi), \\
		s_k(\xi) &= s_{k-1}'(\xi) + s_0(\xi) \mathcal{R}_{k-1}(\xi),
		\label{aimrecurrence}
	\end{align}
	with $\mathcal{R}_{-1} = 0$, $s_{-1} = 1$, and $k = 0,1,2,\dots$. The quantization condition is given by
	\begin{align}
		\delta_k(\xi) = \mathcal{R}_k(\xi) s_{k-1}(\xi) - \mathcal{R}_{k-1}(\xi) s_k(\xi) = 0.
		\label{aimquantization}
	\end{align}
	Solving $\delta_k(\xi_0)=0$ at $\xi_0 = 1 - r_s/r_0$ yields the QNM frequencies $\omega_{\ell n}$.

\begin{figure}
	\centering
	\includegraphics[width=1.0\linewidth]{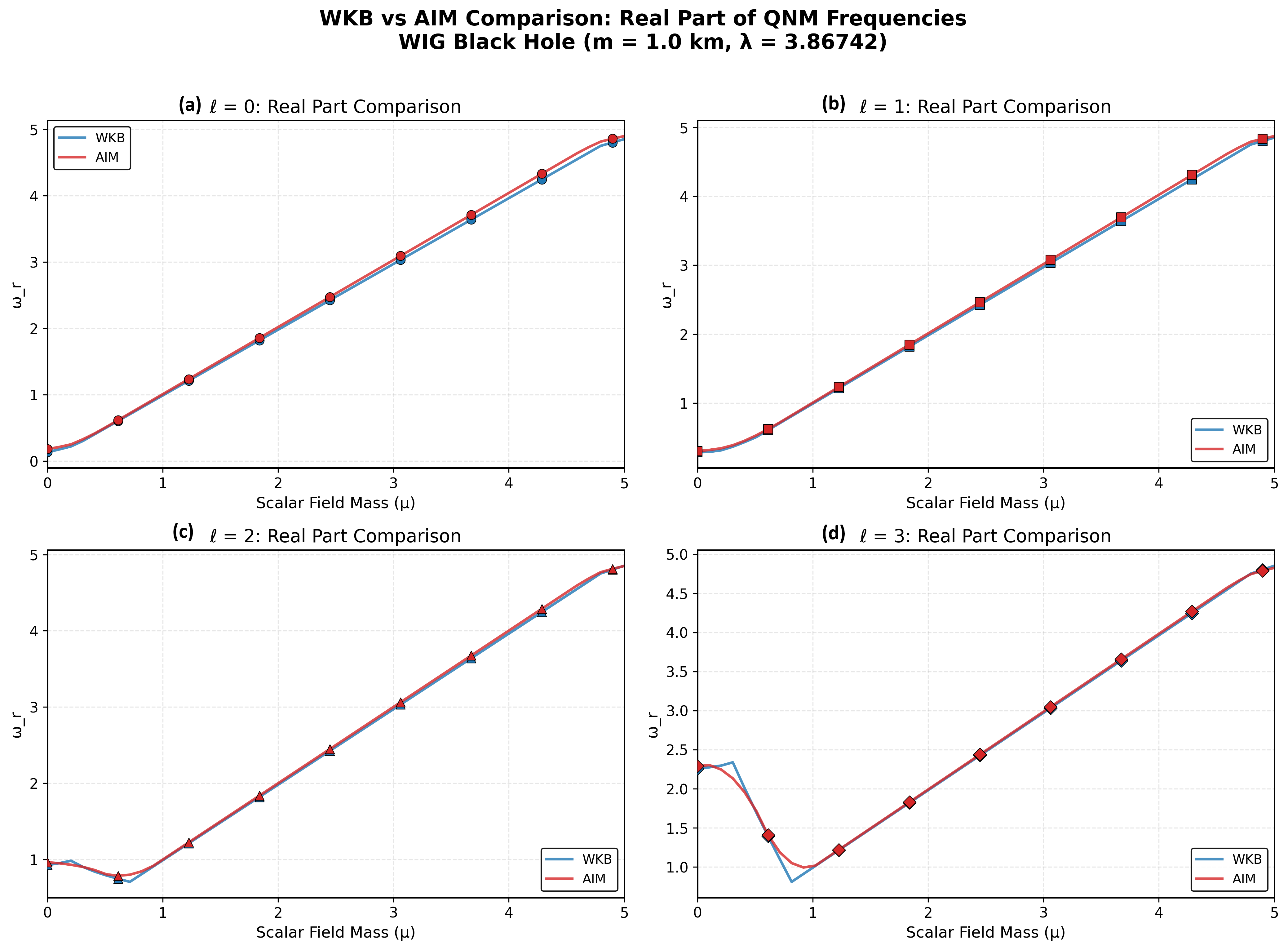}
	\caption{Comparison of the real parts of quasi-normal mode frequencies ($\omega_r$) computed using 
		WKB (blue lines) and AIM (red lines) methods for massive scalar fields in a WIG black hole 
		with mass $m = 1$ km and Weyl coupling $\lambda = 3.869$. The panels show different 
		angular momentum quantum numbers: (a) $\ell = 0$, (b) $\ell = 1$, (c) $\ell = 2$, and 
		(d) $\ell = 3$. The real part $\omega_r$ represents the oscillation frequency of the 
		perturbation modes. Both methods show $\omega_r$ increasing monotonically with the scalar 
		field mass $\mu$, indicating that heavier scalar fields oscillate at higher frequencies. 
		The markers on the curves indicate sampled data points for 
		clarity.}
	\label{realqnm}
\end{figure}

\begin{figure}
	\centering
	\includegraphics[width=1.0\linewidth]{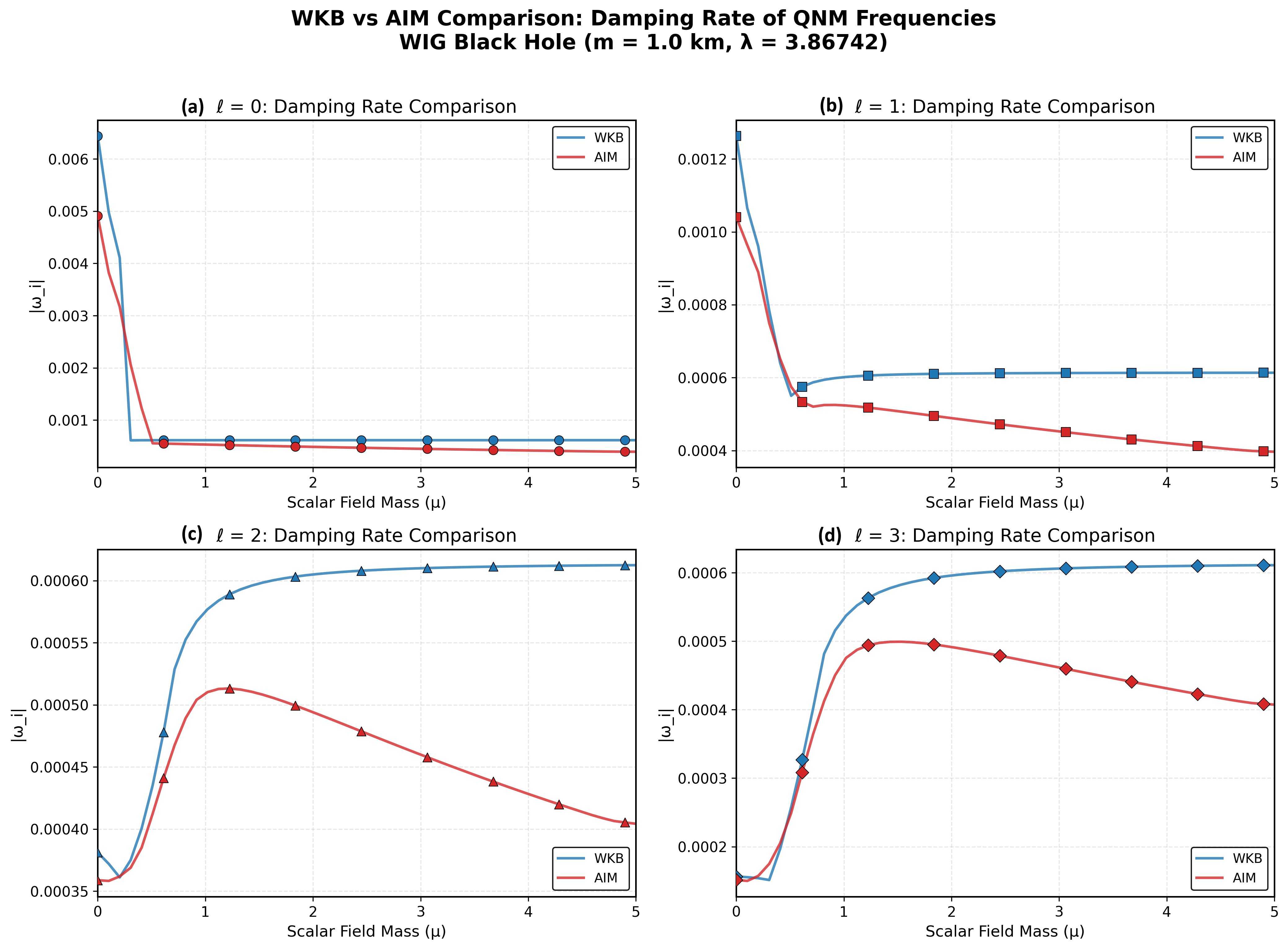}
	\caption{Comparison of the damping rates ($|\omega_i|$) computed using WKB (blue lines) and AIM 
		(red lines) methods for massive scalar fields in a WIG black hole with mass $m = 1$ km 
		and Weyl coupling $\lambda = 3.869$. The panels show different angular momentum quantum 
		numbers: (a) $\ell = 0$, (b) $\ell = 1$, (c) $\ell = 2$, and (d) $\ell = 3$. The 
		damping rate $|\omega_i|$ represents the inverse of the decay time scale of the 
		perturbation modes. Both methods show that $|\omega_i|$ generally decreases with 
		increasing scalar field mass $\mu$, indicating that heavier scalar fields produce 
		longer-lived perturbations. The markers on the curves indicate sampled 
		data points for clarity.}
	\label{imageqnm}
\end{figure}

The obtained results of the QNM frequencies from WKB and AIM analysis are shown in Fig. \ref{realqnm} for the case of real frequencies, and in Fig. \ref{imageqnm} for the case of imaginary frequencies. Our analysis reveals that all computed modes have negative imaginary parts (\(\omega_i < 0\)) for the configurations considered, indicating exponential decay of scalar perturbations within the test-field approximation. The ratio \(\omega_r/|\omega_i|\), which corresponds to the quality factor of the oscillation, reaches values as high as \(\sim 10^4\), indicating that these perturbations undergo thousands of cycles before decay. The results from the 6th-order WKB approximation \cite{Konoplya2003} and the AIM \cite{Cho2010} are in agreement for these long-lived modes, typically showing a relative discrepancy of less than \(1\%\) in the real part \(\omega_r\) for \(\mu > 1\). However, for the fundamental mode \((\ell=0, \mu=0)\), the methods show a more significant variance, with a relative error of about \(30\%\) in \(\omega_r\). This is consistent with the known limitation of the WKB method for low \(\ell\) values and highlights the advantage of using complementary numerical techniques like AIM to verify results in these regimes. Our findings of long-lived scalar modes with \(\omega_i < 0\) in the WIG spacetime align with and extend previous analyses in conformal gravity spacetimes \cite{Momennia2018}, reinforcing the notion that modifications to the gravitational Lagrangian can have profound implications for the ringdown phase and the resultant gravitational wave signatures from perturbed black holes.
\subsection{The Time Domain Integration Method \label{timedomainintegration}}
\begin{figure}
	\centering
	\includegraphics[width=1.0\linewidth]{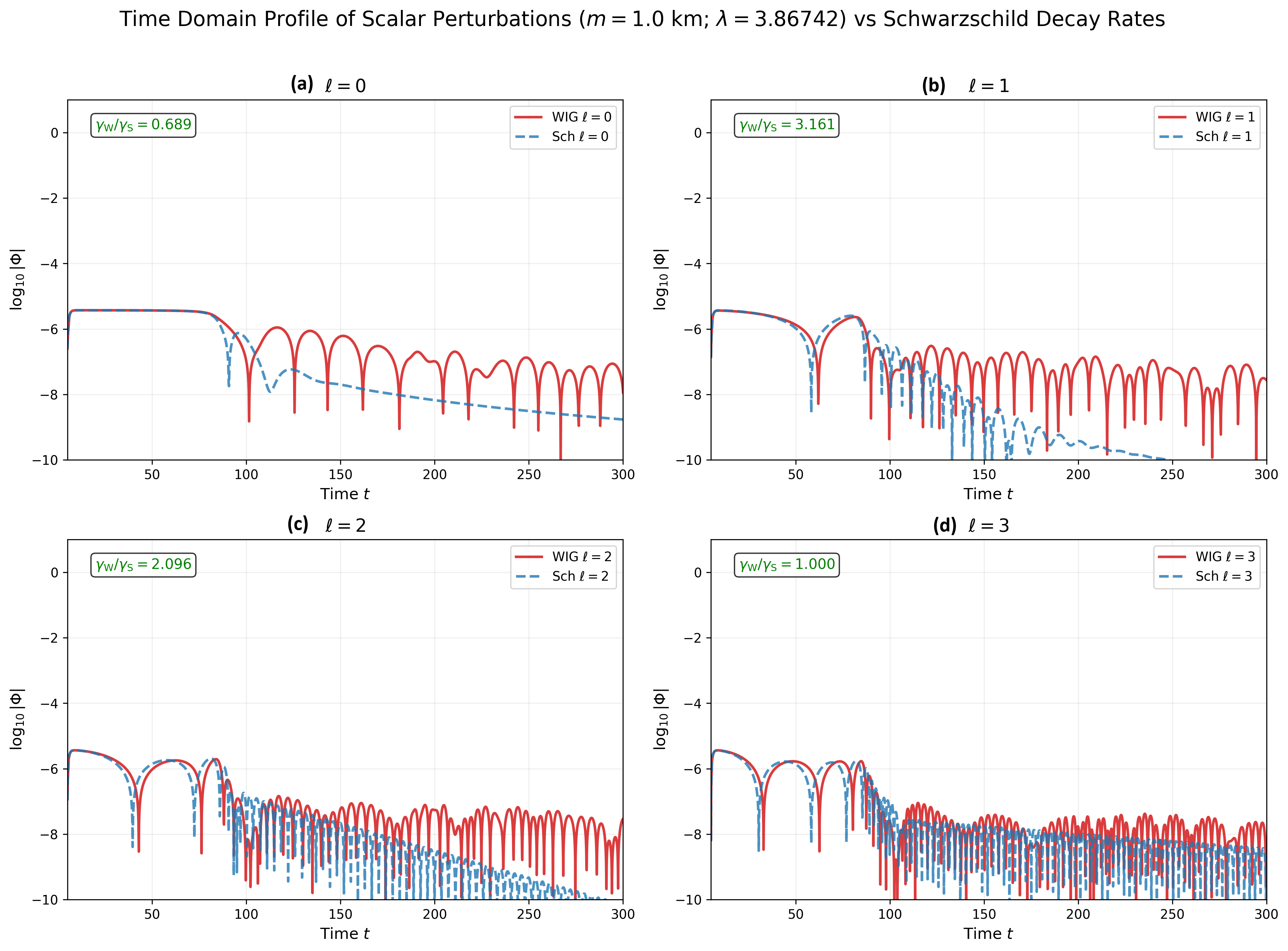}
	\caption{Time-domain profiles of scalar field perturbations for multipoles $\ell=0,1,2,3$ with $m=1.0$ km and $\lambda=3.869$. The WIG background (red, solid) is compared with the corresponding Schwarzschild case (blue, dashed).}
	\label{timedomainanalysis}
\end{figure}

To study the time evolution of perturbations, we transform eq. (\ref{masterequation}) to characteristic null coordinates $u = t - r_*$ and $v = t + r_*$, yielding 
\begin{align}
\frac{\partial^2 \Phi}{\partial u \partial v} + \frac{1}{4} V_{\ell}(r) \Phi = 0.
\label{nullwave}
\end{align}
This two-dimensional wave equation is particularly amenable to numerical integration using the finite difference method. We discretize the $(u,v)$ plane with step sizes $\delta u = \delta v = h$, and employ the following finite difference scheme
\begin{align}
&\Phi(u+h, v+h) = \Phi(u, v+h) + \Phi(u+h, v) - \Phi(u, v) \nonumber \\
&- \frac{h^2}{8} V\left(r_* = \frac{2v - 2u + h - h}{4}\right) \left[\phi(u+h, v) + \Phi(u, v+h)\right] + \mathcal{O}(h^4).
\label{finitediff}
\end{align}
The initial perturbation is modeled as a Gaussian pulse in the advanced time coordinate $v$
\begin{align}
\Phi(u = u_0, v) = \exp\left[-\frac{(v - v_c)^2}{2\sigma^2}\right],
\label{initialgaussian}
\end{align}
where $v_c = 50\,m$ determining the peak location and $\sigma = 2m$ controlling the width. The initial data on the retarded time slice $v = v_0$ is set to $\phi(u, v_0) = 0$, corresponding to no incoming radiation from past null infinity. The numerical integration is performed on a rectangular grid in the $(u,v)$-plane, with $u, v \in [0, 1000m]$. To ensure numerical stability, we employ a Courant factor of $0.5$ and implement fourth-order Runge-Kutta time stepping with adaptive step-size control. Fig. \ref{timedomainanalysis}  shows the time domain evolution of scalar perturbations for multipole indices $\ell = 0,1,2,3$ in the WIG spacetime, compared with the corresponding Schwarzschild background. At early times, the waveforms nearly coincide for all $\ell$, indicating that the prompt response is largely insensitive to the underlying spacetime geometry. Differences emerge in the intermediate and late--time regimes, where the Schwarzschild perturbations exhibit faster damping, while the WIG case shows prolonged quasinormal ringing and a slower decay. This effect is particularly pronounced for low and intermediate multipoles ($\ell = 1,2$), as reflected by the larger ratios $\gamma_{\rm WIG}/\gamma_{\rm Sch}$, whereas higher multipoles ($\ell = 3$) display decay rates closer to the Schwarzschild limit. These results suggest that modifications in the effective potential of the WIG spacetime enhance wave trapping and back--scattering, leading to long--lived scalar perturbations relative to the Schwarzschild case.
\section{Results and Discussion \label{results}}

Black hole shadows offer a unique window to study modifications to standard GR in the strong-field regime. Similarly, the QNM spectrum emitted during the ringdown of an astrophysical black hole provides a complementary probe of the spacetime geometry. In this paper we have studied MORD \cite{lee2019motion}, referred to here as WIG. We estimated the shadow parameters for different black hole masses in a spherically symmetric static spacetime. As a first approximation we assumed that these black holes are surrounded by a constant-density baryonic matter halo, and we also added a homogeneous plasma background to check the deviation in the coupling constant $\lambda$.

We stress that the shadow analysis presented here is a theoretical exploration under simplified assumptions. As discussed in Refs. \cite{Psaltis2015, Gralla2021, Volkel2021, Kostas2021, Guillermo2021, Bauer2022, Prashant2022, Dimitry2022}, there exist degeneracies between modified gravity parameters and unknown astrophysical processes, for example, the geometry and emission profile of the accretion flow. A more robust analysis would require a more realistic astrophysical model and is left for future work.

The shadow radii obtained in Table \ref{photonsphereandshadowradius} represent theoretical predictions of the WIG framework under the specific assumption of a constant-density halo with $\rho_c = 0.5$. They are not intended as observational constraints, as a direct comparison with EHT data would require solving the theory at the Sgr A* mass scale with a realistic halo profile. With the addition of a homogeneous plasma, the shadow size reduces compared to the case with only the baryonic matter halo; higher plasma density, $\rho_p = 0.4$, produces smaller shadows than $\rho_p = 0.2$.

The QNM analysis reveals that all computed modes have negative imaginary parts (\(\omega_i < 0\)) for the configurations considered, indicating exponential decay of scalar perturbations in the test-field approximation. As shown in Figs.~\ref{realqnm}--\ref{imageqnm}, for field masses \(\mu > 0\) we find quality factors \(Q \equiv \omega_r/|\omega_i|\) reaching \(\mathcal{O}(10^3-10^4)\), meaning that these perturbations oscillate for thousands of cycles before decaying. This phenomenon arises from the effective potential modifications induced by the Weyl-matter coupling and could, in principle, serve as a signature in future gravitational-wave observations. However, we emphasize that our QNM calculation is performed in the test-field approximation. A true gravitational QNM calculation would require perturbing the metric, the matter tensor, and the interaction term $I_{\mu\nu}$ simultaneously. The time-domain integration (Fig.~\ref{timedomainanalysis}) shows exponential decay of the perturbations with no evidence of growing modes for the scalar perturbations considered in the test-field approximation, with the decay following characteristic quasinormal ringing and power-law tail behaviour \cite{Chandrasekhar1983}.

The results obtained from the WIG framework demonstrate that the theory can produce deviations from GR in the strong-coupling regime. These deviations are parameterized by the coupling constant $\lambda$, which, under our specific assumptions, takes values $\lambda \sim 3.5$--$3.9$. In scalar-tensor-vector gravity (MOG), analyses of rotating black holes have constrained the MOG parameter to $0.350 \lesssim \alpha_{\text{up}} \lesssim 0.485$ for M87* and $0.162 \lesssim \alpha_{\text{up}} \lesssim 0.285$ for Sgr A* \cite{Khodadadi2022}. Yukawa gravity modifications have been tightly constrained using both M87* and Sgr A* shadows, yielding coupling parameters $\kappa = -0.01 \pm 0.17$ for M87* and $\kappa = -0.04^{+0.09}_{-0.10}$ (Keck prior) or $\kappa = -0.08^{+0.09}_{-0.06}$ (VLTI prior) for Sgr A* at large Yukawa wavelengths, with no significant deviation from GR detected \cite{Tan2025a}. In Weyl-Cartan theory, Kerr-Newman black holes have been evaluated against EHT shadow estimates, demonstrating that such frameworks can be distinguished from GR through precision shadow measurements \cite{Jafarzade2023}. Fourth-order conformal Weyl gravity similarly predicts shadow morphologies that remain largely indistinguishable from GR unless the conformal parameters $\gamma$ and $\kappa$ are increased by several orders of magnitude beyond values already ruled out by gravitational phenomenology \cite{Nelson2017}. Within this comparative landscape, our WIG results show that the coupling parameter produces shadow radii that differ from the GR prediction under the specific assumptions of our model. This highlights the importance of future work with realistic halo profiles to determine whether WIG can be distinguished from GR observationally. The exceptionally long-lived QNMs we find for massive scalar fields ($Q \sim 10^3-10^4$) provide an interesting phenomenological prediction that, combined with further theoretical studies, positions WIG as an interesting modified gravity framework worthy of further theoretical study, particularly with more realistic halo models and inclusion of black hole rotation.

\section*{Acknowledgments}
 We would like to acknowledge Prof. Asghar Qadir for his suggestion on the work. We are grateful to the anonymous referees for their constructive comments, which have significantly improved the quality and clarity of this work.

\end{document}